# Superconducting Nanoelectromechanical Transducer Resilient to Magnetic Fields

Jinwoong Cha[1], Hak-Seong Kim[1], Jihwan Kim[1,2], Seung-Bo Shim[1], and Junho Suh[1,*]

[1]Quantum Technology Institute, Korea Research Institute of Standards and Science, Daejeon, South Korea

[2]Department of Physics, Korea Advanced Institute of Science and Technology, Daejeon, South Korea

Correspondence to Junho Suh: junho.suh@kriss.re.kr

## Abstract

Nanoscale electromechanical coupling provides a unique route towards control of mechanical motions and microwave fields in superconducting cavity electromechanical devices. Though their successes in utilizing the optomechanical or electromechanical back-action effects for various purposes, aluminum imposes severe constraints on their operating conditions with its low superconducting critical temperature (1.2 K) and magnetic field (0.01 T). To extend the potential of the devices, here we fabricate a superconducting electromechanical device employing niobium and demonstrate a set of cavity electromechanical dynamics including back-action cooling and amplification, and electromechanically induced reflection at 4.2 K and in strong magnetic fields up to 0.8 T. This device could be used to realize electromechanical microwave components for quantum technologies by integrating amplifiers, converters, and circulators on a single chip that can be installed at the 4K stage of dilution refrigerators. Moreover, with its ability to control and readout nanomechanical motions simultaneously, this niobium electromechanical transducer could provide powerful nanomechanical sensing platforms.

**Introduction**

The motion of a nanoscale mechanical object implemented in an electromagnetic cavity at microwave[1,2] (~ GHz) or optical frequencies[3-5] (> 100 THz) can be detected by measuring the motion-induced intensity modulation of the electromagnetic field coming out from the cavity. This interferometric detection method is essential in measuring nanomechanical motions as it enables picometer-level sensitivity. Thanks to this powerful method, nanomechanical devices are considered promising for sensing small perturbations like mass-loading by single molecules[6-8]. More sensitive detection requires the enhancement of the quality factor, $Q$, of the optical cavity, which ultimately leads to strong interactions of mechanical and optical eigenmodes (i.e. phonons and photons) via the radiation pressure, enabling energy transfer between different physical domains[9]. This dynamic process is, in general, described by a linearized optomechanical Hamiltonian[10], $\hat{H} = -\hbar \Delta \hat{a}^\dagger \hat{a} + \hbar \Omega_m \hat{b}^\dagger \hat{b} + \hat{H}_{int}$ in the rotating frame at $\omega_d$, where $\Delta = \omega_d - \omega_c$ is the detuning frequency, $\omega_d$ is the driving frequency, $\omega_c$ is the resonant frequency of an optical cavity, $\Omega_m$ is the resonant frequency of a mechanical resonator. $\hat{a}(\hat{a}^\dagger)$ and $\hat{b}(\hat{b}^\dagger)$ are the annihilation (creation) operators for photon and phonon modes, respectively, and $\hat{H}_{int}$ is the photon-phonon interaction Hamiltonian. If the system is driven in the red-detuned regime ($\Delta = -\Omega_m$), $\hat{H}_{int} = \hbar g(\hat{a}^\dagger \hat{b} + \hat{a}\hat{b}^\dagger)$, which describes coherent exchange of the two bosonic modes, resulting in the cooling of mechanical motions. In contrast, a blue-detuned drive ($\Delta = \Omega_m$) sets $\hat{H}_{int} = \hbar g(\hat{a}^\dagger \hat{b}^\dagger + \hat{a}\hat{b})$, which leads to parametric down-conversion of the drive photons, increasing phonons in the mechanical resonator and photons in the optical cavity. In both cases, $g = g_0\sqrt{n_d}$ denotes the optomechanical coupling rate when the number of the drive photons in the cavity is $n_d$. $g_o = \frac{\partial \omega_c}{\partial x}\sqrt{\frac{\hbar}{2m_{eff}\Omega_m}}$ is the single photon coupling rate, where $m_{eff}$ is the effective mass of the mechanical resonator.

Cavity optomechanics have been explored in various experimental platforms such as nanomembranes in the middle of high $Q$ optical cavities[11], toroidal optical cavities[12,13], optomechanical crystals[14-16], and superconducting electromechanical devices[17-27]. Among these, the superconducting electromechanical devices provide a compact, on-chip scale platform without any complicated optical alignments, offering unprecedented opportunities for manipulating both nanomechanical motions and microwave signals via strong electromechanical coupling at the nanoscale. This coupling leads to the parametric interaction of nanomechanical resonance modes and superconducting microwave cavities, enabling their mutual energy exchange. Such electromechanical devices have recently shown their operations in the quantum regime, demonstrating cooling[17-19], squeezing[20-21], and amplification[22] of mechanical motions, photon-phonon entanglement[23], electromechanically induced transparency[24,25], and nonreciprocal microwave transmission[26,27]. Though these examples show their great promise, the conditions for their reliable operations are limited by superconducting critical temperature (1.2 K) and magnetic field (0.01 T) of aluminum. Extending the operating conditions of the devices is strongly desired to utilize their versatility as nanomechanical sensors and microwave components for quantum technologies.

Here we describe the realization of a superconducting nanoelectromechanical transducer with niobium and demonstrate their operations at 4.2 K and in strong magnetic fields (up to 0.8 T). We show that the micro-fabricated niobium microwave resonator exhibits relatively high quality factor (~ 4,000) at 4.2 K and preserves its microwave resonance even in external magnetic fields. We demonstrate efficient back-action cooling and amplification of nanomechanical motions, and electromechanically induced reflection of microwave signals.

We also experimentally show the detection of nanomechanical motions in 0.8 T magnetic field, manifesting the potential for exploring spin-phonon interactions[28-31] using the niobium device.

**Results**

**Niobium superconducting electromechanical transducer.** We explore the optomechanical interaction of microwave photons and nanomechanical phonons by fabricating a *LC* microwave resonator circuit made of superconducting niobium (Fig. 1a) where a free-standing, 130-nm thick nanomechanical resonator forms a vacuum-gap capacitor with a 100-nm thick bottom electrode (Fig. 1b). This membrane resonator is connected to a spiral inductor and the bottom electrode is connected to the circuit ground. The diameter of the membrane is 12.6 μm and the bottom electrode is 10 μm in diameter. The effective mass, $m_{eff}$, of the membrane is ~ 140 pg based on its dimensions. Our device is a typical set-up for studying microwave cavity electromechanics[19-21,23,24,32] via the dispersive electromechanical coupling, which induces the shift of the microwave resonant frequency by the motion of the membrane. The nanomechanical resonator is suspended over the bottom electrode with a 130-nm vacuum gap by releasing a sacrificial layer of Poly-methyl-methacrylate (PMMA) (Fig. 1c and Methods). To characterize microwave and mechanical responses of our device, we perform all the measurements at 4.2 K, the liquid helium temperature (Fig. 1d). The microwave resonator is capacitively coupled to the input and the output ports to drive and measure the responses of the device. From a microwave transmission measured with a network analyzer, we obtain $\omega_c$ (≈ 2π × 3.777 GHz) and the total decay rate of the microwave resonator, $\kappa$ (≈ $\kappa_o$ + $\kappa_{ext}$ = 2π × 960 kHz) with intrinsic loss $\kappa_o$ (≈ 2π × 330 kHz) and external loss $\kappa_{ext}$ (≈ 2π × 640 kHz) (Fig. 1e). We characterize the mechanical responses by measuring a noise spectrum around $\omega_c$. This spectrum results from the frequency up-conversion of the thermomechanical noise around $\Omega_m$ when the device is excited by a red-detuned microwave drive at $\omega_d = \omega_c - \Omega_m$. This microwave drive is weak enough to induce negligible optomechanical back-action effects. We also obtain $\Omega_m$ ($\approx 2\pi \times 8.4$ MHz) and the intrinsic mechanical damping rate, $\Gamma_m$ ($\approx 2\pi \times 810$ kHz) (Fig. 1f). Note that this $\Omega_m$ corresponds to the fundamental flexural mode of the membrane as the bottom electrode provide an isotropic electrostatic force distribution. The ratio of $\Omega_m/\kappa \sim 9$ implies that our system is in the sideband resolved regime, meaning that only one of the two sidebands (Stokes or anti-Stokes) is dominant depending on $\Delta$.

**Cooling, amplification, and detection of nanomechanical motions.** The parametric coupling of microwave photons and nanomechanical phonons enables simultaneous control and measurement of the dynamics of a nanomechanical resonator. To investigate the optomechanically induced behavior of our nanomechanical resonator, we excite our device in the red-detuned (blue-detuned) regime and measure total mechanical displacement noise spectrum, $S_x(\omega)$, of the up-converted (down-converted) sideband signals. The equipartition theorem for an harmonic oscillator gives $<x^2(t)> = k_B T_m /(m_{eff}\Omega_m^2) = \int_{-\infty}^{\infty} S_x(\omega)d\omega/2\pi$, where $<x^2(t)>$ is the variance of the mechanical displacement, and $T_m$ is the temperature of the vibration mode[10]. Therefore, $S_x(\omega)$ reveals the thermal energy of the mechanical resonator (see Methods for detailed discussion). To see the effects of the dynamic back-action in our electromechanical device, we vary the detuning frequency, $\Delta = -\Omega_m + \delta$, from $\delta = -1$ to 1 MHz (Supplementary Fig. 1). We observe the consequent optomechanical spring and damping effects which originates from the retardation effect in the cavity field[10]. The optomechanical damping clearly shows its maximum at $\Delta = -\Omega_m$. We also characterize the back-action cooling and amplification performance of our device at $\Delta = -\Omega_m$ (Fig. 2a) and $\Delta = \Omega_m$ (Fig. 2b) for different drive powers, $n_d$. For $\Delta = -\Omega_m$, $S_x(\omega)$ of the up-converted sidebands, which exhibit Lorentzian resonance curves, show significant linewidth broadening, decreasing the amplitude of the resonance peaks as we increase $n_d$ (Fig. 2a). The signal-to-noise ratio (SNR), $S_x^{mech}(\Omega_m)/S_x^{bg}(\Omega_m)$ also changes with increasing $n_d$, manifesting the interplay of the thermomechanical noise, the measurement imprecision and the dynamic back-action (Supplementary Fig. 2a). We obtain the best force sensitivity of 400 aN/Hz$^{1/2}$ at $n_d \approx 4 \times 10^6$, which is limited by the thermomechanical noise (Supplementary Fig. 2b). On the other hand, down-converted $S_x(\omega)$ at $\Delta = \Omega_m$ shows linewidth narrowing with increasing peak amplitudes, indicating the enhancement of quality factor $Q$ (Fig. 2b). We perform quantitative analyses of $S_x(\omega)$ and extract phonon numbers of our mechanical resonator (see Methods). The thermal occupation factor of the nanomechanical mode at $T_{bath}$ = 4.17 K is $n_{ph}$ = $[\exp(\hbar\Omega_m/k_BT_{bath}) - 1]^{-1} \approx 1.04 \times 10^4$. In the red-detuned regime ($\Delta = -\Omega_m$), we obtain $n_{ph} \approx$ 188 at the maximum drive power ($n_d \approx 6.5 \times 10^8$) (Fig. 2c). This $n_{ph}$ corresponds to 76 mK in terms of mode temperature $T_m$. We also extract the single-photon coupling rate, $g_0$ ($\approx 2\pi \times 3.3$ Hz), from this cooling experiment (see Methods). Though this value is below the one obtained from aluminum-based electromechanical systems[19], it is large enough to enable back-action effects. The coupling can be further enhanced by increasing the microwave resonant frequency and decreasing the vacuum gap distance. For a blue-detuned drive ($\Delta = \Omega_m$), the phonon number increases to $n_{ph} \approx 3.8 \times 10^4$ ($T_m \approx 15.4$ K) at the largest drive power ($n_d \approx 5.2 \times 10^6$) (Fig. 2c). We also note that the damping rate of the mechanical resonator, $\Gamma_T$, in this blue-detuned regime is diminished to $2\pi \times 214$ Hz, indicating four-fold enhancement of $Q$ (Fig. 2d). The ability to tune $Q$ and measure the nanomechanical motion simultaneously is of unique aspect and could be harnessed in nanomechanical sensors to achieve enhanced SNR[33].

**Electromechanically Induced Reflection.** Engineering the transmission of microwave signals is also another distinct feature of superconducting electromechanical devices[24-27]. A related phenomenon to this is the electromechanically induced transparency[24,25] (EMIT), which is analogous to the electromagnetically induced transparency found in atom-based cavity quantum electrodynamics[34]. EMIT leads to a transparency window for the probe signals near the microwave resonant frequency ($\omega_c$). The probe signal in addition to the red-detuned drive ($\Delta = -\Omega_m$) resonantly excites the nanomechanical motion. This driven nanomechanical motion induces the anti-Stokes sideband of the red-detuned drive around the microwave resonant frequency, resulting in destructive interference with the probe signal and the onset of the transmission window. In our device, the two-port geometry leads to electromechanically induced reflection (EMIR) which denote a reflection window in the transmission resonance (Fig. 3a). In the resolved sideband limit, the transmission spectrum of the probe signal for a two-port electromechanical device is formulated by considering the Heisenberg picture of the linearized optomechanical Hamiltonian in the rotating frame at $\omega_d$[12,21]. The equation for the transmission $\mathcal{T}$ is given by,

$$\mathcal{T} = \frac{\kappa_R \kappa_L}{\left(\frac{\kappa}{2}\right)^2 + (\Delta_p - \Omega_m)^2 + \frac{g_0^2 n_d \left\{g_0^2 n_d + \frac{\kappa \Gamma_m}{2} - 2(\Delta_p - \Omega_m)^2\right\}}{\left(\frac{\Gamma_m}{2}\right)^2 + (\Delta_p - \Omega_m)^2}}.$$

Here, $\Delta_p = \omega_p - \omega_d$ is the probe detuning. $\kappa_R$ and $\kappa_L$ denote the input and output couplings and equal to $\kappa_{ext}/2$. We learn from this equation that the linewidth and depth of the reflection window strongly depends on the drive power, $n_d$. To systematically study the EMIR effect in our device, we measure the transmission spectrum of a weak probe signal at $\omega_p$ with the red-detuned drive ($\Delta = -\Omega_m$) while we vary $n_d$ from $2 \times 10^4$ to $6.5 \times 10^8$ (Fig. 3b). When $\omega_p \approx \omega_c$, the depth of the reflection window, $A_{EMIR}$, increases as $n_d$ increases and reaches −32 dB at the maximum drive power ($n_d \approx 6.5 \times 10^8$). The linewidth of the reflection window, $\Gamma_{EMIR}$, corresponds to the damping rate of our nanomechanical resonator and broadens up to 32 kHz at the maximum $n_d$ (Fig. 3c). In the resolved sideband regime, the window linewidth at $\Delta = -\Omega_m$ is given by $\Gamma_{EMIR} = \Gamma_m(1+C)$, where $C = 4g_0^2 n_d/(\Gamma_m \kappa)$ represents the electromechanical cooperativity. In our device, we obtain the maximum cooperativity of $C \approx 40$. Note that $C \geq 1$ leads to notable electromechanical back-action effects[12]. Based on the currently available $g_0$ ($\approx 2\pi \times 150$ Hz) demonstrated in aluminum-based devices[24], it should be possible to engineer our niobium device to achieve $C \approx 10^5$ at 4 K.

**Electromechanical transduction in strong external magnetic fields.** Niobium is also expected to enable reliable operation of superconducting electromechanical devices in magnetic fields, which could enable studying and harnessing, for example, spin-phonon interactions[28-31] in hybrid quantum devices. To support this idea, we study the responses of our electromechanical device to an external magnetic field, $B$. We apply the field in the direction parallel to the device and measure microwave transmission while varying $B$ from 0 to 0.89 T, with 0.089 T steps (Fig. 4a). We find that $\omega_c$ decreases from 3.778 GHz (B = 0 T) to 3.744 GHz (B = 0.8 T) and $\kappa$ increases from 950 kHz (B = 0 T) to 31 MHz (B = 0.8 T) (Fig. 4b). This increase of $\kappa$ mainly originates from quasiparticle generation in niobium[35]. $\kappa$ becomes comparable to $\Omega_m$ around 0.5 T and starts to exceed $\Omega_m$. In this regime, the resolved sideband approximation becomes irrelevant as the contribution of the Stokes sideband is not negligible. To characterize electromechanical responses, we measure the EMIR spectra for different $B$ by driving the device in the red-detuned regime ($\Delta = -\Omega_m$). The highest $B$ at which we observe the EMIR effect is 0.8 T with $A_{EMIR}$ of ~ 0.5 dB at $n_d \approx 3.35 \times 10^7$ (Fig. 4c). Even at $B$ = 0.8 T, we still observe non-negligible electromechanical back-action damping effect with $C$ ~ 0.46. $\Gamma_{EMIR}$ exhibits reduced electromechanical effects as $B$ increases (Fig. 4d). This mainly originates from the decrease of the single-photon electromechanical cooperativity, $C_0 = 4g_0^2/(\Gamma_m \kappa)$, due to the increase of $\kappa$. To check this, we estimate $g_0$ from $C_0$ for different $B$ and it remains almost constant up to 0.62 T and starts to decrease as the resolved sideband approximation becomes invalid (Fig. 4d inset).

**Discussion**

Utilizing the cavity-electromechanical transduction scheme could be a promising approach for nanomechanical spin-sensing[36-38] and quantum transduction for long-distance spin-spin interactions[39,40]. A typical electromechanical Hamiltonian considering spin-phonon interactions reads $\widehat{H} = -\hbar \Delta \hat{a}^\dagger \hat{a} + \hbar\Omega_m \hat{b}^\dagger \hat{b} + \hbar g \hat{a}^\dagger \hat{a}(\hat{b}^\dagger + \hat{b}) + \sum_i \hbar \lambda_i [\hat{b}^\dagger \hat{\sigma}_{-,i} + \hat{b} \hat{\sigma}_{+,i}]$. Here, $\lambda_i$ is the spin-phonon coupling rate of a nanomechanical mode and the $i^{th}$ spin and $\hat{\sigma}_{+,i}$ and $\hat{\sigma}_{-,i}$ are the raising and the lowering operators for the spin. Inferring from this Hamiltonian, the electromechanical transduction methods such as back-action cooling, amplification, and squeezing could be employed to control and detect spins of interest. Moreover, interfacing spins at different locations could be possible via microwave photons or optical light with the aid of electromechanical or optomechanical transductions. The resonant frequency of our mechanical resonator (~ 8 MHz) matches well with the nuclear magnetic resonances of species like $^1$H ($\gamma$ ~ 42.57 MHz/T, where $\gamma$ is a gyromagnetic ratio)[36]. A strategy to couple spins and nanomechanical motions could be to implement a magnetic tip made of a ferromagnetic material like NdFeB[38] on the top of the nanomechanical membrane and position the tip in the vicinity of the spins. Thanks to the superior mechanical properties of niobium, the magnetic tip of hundreds of nanometers in size should not significantly modify the mechanical properties of the membrane and also the performance of the electromechanical transduction.

For microwave engineering in quantum technologies, this niobium device could take advantage of the novel principles discovered in multimode electromechanical device. For example, routing and amplifying microwave signals in non-reciprocal ways[26,27] could be achieved in a single-chip by integrating various types of electromechanical devices. Especially, their performance stability at 4.2 K supports that the proposed integrated microwave component could offer alternative solutions to the conventional cryogenic microwave isolators and amplifiers by providing compact package size with their small device footprints.

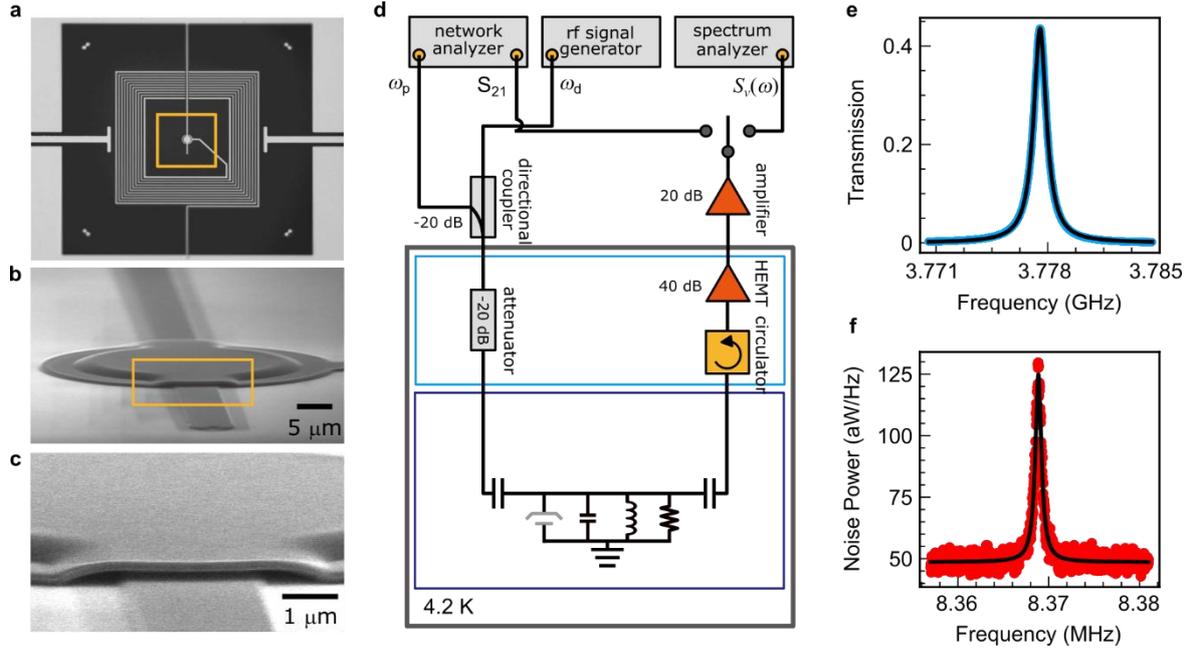

**Fig. 1 Niobium superconducting electromechanical device. a** Optical microscope image of the device. **b** Scanning electron microscope (SEM) image showing the nanomechanical resonator marked by the orange box in (**a**). Scale bar, 5 μm. **c** A magnified SEM image of the vacuum gap region marked by the orange box in (**b**). Scale bar, 1 μm. **d** Schematic of the microwave measurement setup. The network analyzer is used to measure transmission spectra ($S_{21}$) of the probe signal at $\omega_p$. The rf signal generator sends the microwave drive signal at $\omega_d$. The spectrum analyzer is used to measure noise spectrum $S_v(\omega)$ in units of W/Hz. The circuit components enclosed in the dark-blue box describe the electromechanical device. **e** Transmission spectrum of the microwave resonator (blue) measured by the network analyzer. The black solid line represents a Lorentzian fit of the microwave resonance. **f** Thermomechanical noise spectrum (red) of the nanomechanical motion at 4.17 K. The resonator is driven by a weak red-detuned drive to avoid the back-action damping. The black solid line is a Lorentzian fit of the mechanical resonance.

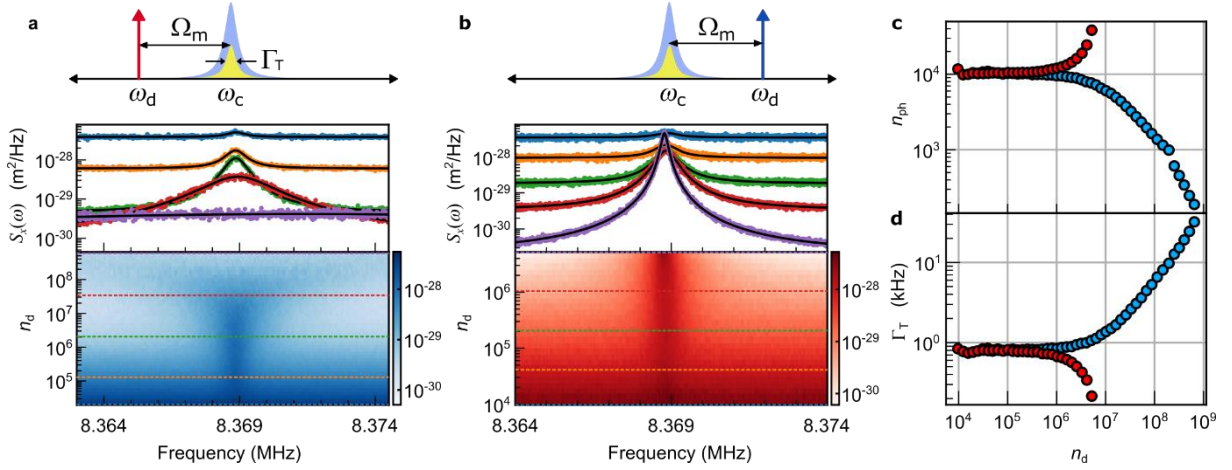

**Fig. 2 Sideband cooling and amplification of the mechanical motion. a, b** Schematic describing the experiments in frequency domains. The device is driven by **(a)** red-detuned ($\Delta = -\Omega_m$) and **(b)** blue-detuned ($\Delta = \Omega_m$) microwave signals with different drive powers. The blue shaded Lorentzian curves indicate the microwave cavity mode. The yellow Lorentzian curves show the sideband generated by the detuned microwave drives. The plots and the colormaps show the mechanical displacement spectral density, $S_x(\omega)$, for different drive powers, $n_d$. The color bars next to the colormaps represent the corresponding $S_x(\omega)$. The color of the line plots represent $S_x(\omega)$ for different $n_d$ (marked with dashed lines in the colormap). **c** Phonon occupancy $n_{ph}$ as a function of $n_d$. **d** The total mechanical damping rate, $\Gamma_T$, as a function of $n_d$. The blue and red dots in **(c)** and **(d)** correspond to cooling and amplification cases, respectively.

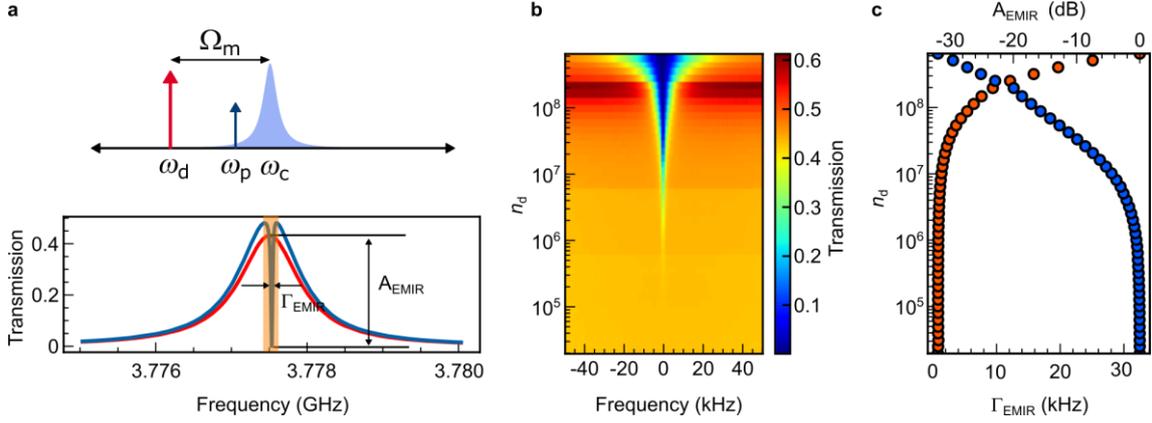

**Fig. 3 Electromechanically Induced Reflection. a** Top: schematic of the experiment in frequency domain. $\omega_d$ is the microwave drive detuned from the cavity resonance by $\Omega_m$. $\omega_p$ is the probe frequency sweeping over the cavity resonance, $\omega_c$. Bottom: microwave resonance spectra for the undriven cavity (red solid line) and the driven cavity (blue solid line). The width and the depth of the reflection window near $\omega_c$ correspond to $\Gamma_{EMIR}$ and $A_{EMIR}$, respectively. **b** Color map showing the transmission spectra near the reflection window marked in (**a**) for different drive powers, $n_d$. **c** $\Gamma_{EMIR}$ (orange dots) and $A_{EMIR}$ (blue dots) as a function of $n_d$.

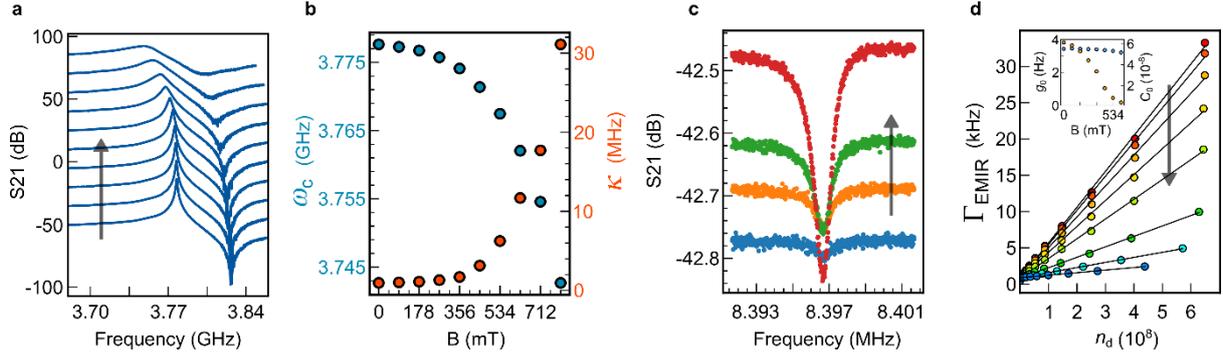

**Fig. 4 Responses of the electromechanical transducer in external magnetic fields. a** Transmission spectrum of the microwave resonator for different external magnetic fields (*B*) ranging from 0 T to 890 mT with 89 mT steps. The arrow denotes the direction of increasing *B*. **b** The resonant frequency (blue dots), $\omega_c$, and the decay rate (orange dots), $\kappa$, of the microwave resonator as a function of *B*. **c** Electromechanically induced reflection at 0.8 T for different drive powers. The arrow indicates increase of the microwave drive power. **d** The linewidth of the EMIR window, $\Gamma_{\text{EMIR}}$, for different *B* ranging from 0 T to 623 mT with 89 mT steps, as a function of the microwave drive, $n_d$. The arrow represents the direction of increasing *B*. The inset shows the single photon coupling rate, $g_0$, (blue dots) and the single photon electromechanical cooperativity, $C_0$, (orange dots) as a function of *B*. To estimate the quantities, we assume the device is operated in the resolved sideband regime.

**References**

1. Regal, C. A., Teufel, J. D., Lehnert & Lehnert, K. W. Measuring nanomechanical motion with a microwave cavity interferometer. *Nat. Phys.* **4,** 555-560 (2008).

2. Teufel, J. D, Donner, T., Castellanos-Beltran, M. A.,Harlow, J. W. & Lehnert, K. W. Nanomechanical motion measured with an imprecision below that at the standard quantum limit. *Nat. Nanotechnol* **4,** 820-823 (2009).

3. Bunch, J. S. *et al*. Electromechanical resonators from graphene sheets. *Science* **315,** 490-493 (2007).

4. Lee, J. *et al*. Electrically tunable single- and few-layer $MoS_2$ nanoelectromechanical systems with broad dynamic range. *Sci. Adv.* **4,** eaao6653 (2018).

5. Unterreithmeier, Q. P., Weig, E. M. & Kotthaus, J. P. Universal transduction scheme for nanomechanical systems based on dielectric forces. *Nature* **458,** 1001-1004 (2009).

6. Yang, Y. T., Callegari, C., Feng, C. X. L., Ekinci, K. L., Roukes, M. L. Zeptogram-scale nanomechanical mass sensing. *Nano Lett.* **6**, 583-586 (2006).

7. Naik, A. K., Hanay, M. S., Hiebert W. K., Feng, X. L. & Roukes, M. L. Towards single-molecule nanomechanical mass spectroscopy. *Nat. Nanotechnol.* **4,** 445-450 (2009).

8. Dominguez-Medina, S. *et al*. Neutral mass spectroscopy of virus capsids above 100 megadaltons with nanomechanical resonators. *Science* **362,** 918-922 (2018).

9. Kippenberg, T. J. & Vahala, K. J. Cavity optomechanics: back-action at the mesoscale. *Science* **321,** 1172-1176 (2008).

10. Aspelmeyer, M., Kippenberg, T. J. & Marquardt, F. Cavity optomechanics. *Rev. Mod. Phys.* **86,** 1391-1452 (2014).

11. Thompson, J. D. *et al*. Strong dispersive coupling of a high-finesse cavity to a micromechanical membrane. *Nature* **452,** 72-75 (2008).

12. Weis, S. *et al*. Optomechanically induced transparency, *Science* **330,** 1520-1523 (2010).

13. Verhagen, E., Deleglise, S., Weis, S., Schliesser, A. & Kippenberg, T. J. Quantum-coherent coupling of a mechanical oscillator to an optical cavity mode. *Nature* **482,** 63-67 (2012).

14. Eichenfield, J., Chan, J., Camacho, R. M., Vahala, K. J. & Painter, O. Optomechanical crystals. *Nature* **462,** 78-82 (2009).

15. Safavi-Naeini, A. H. *et al*. Electromagnetically induced transparency and slow light with optomechanics. *Nature* **472,** 69-73 (2011).

16. Chan, J. *et al*. Laser cooling of a nanomechanical oscillator into its quantum ground state. *Nature* **478,** 89-92 (2011).

17. Teufel, J. D., Harlow, J. W., Regal, C. A. & Lehnert, K. W. Dynamical backaction of microwave fields on a nanomechanical oscillator. *Phys. Rev. Lett.* **101,** 197203 (2008).

18. Rocheleau, T. *et al*. Preparation and detection of a mechanical resonator near the ground state of motion. *Nature* **463,** 72 – 75 (2010).

19. Teufel, J. D. *et al*. Sideband cooling of micromechanical motion to the quantum ground state. *Nature* **475,** 359 – 363 (2011).

20. Pirkkalainen, J.-M., Damskägg, E., Brandt, M., Massel, F. & Sillanpää, M. A. Squeezing of quantum noise of motion in a micromechanical resonator. *Phys. Rev. Lett.* **115,** 243601 (2015).

21. Wollman, E. E. *et al*. Quantum squeezing of motion in a mechanical resonator *Science* **349,** 952 – 955 (2015).

22. F. Massel, T. T. Heikkilä, J.-M. Pirkkalainen, S. U. Cho, H. Saloniemi, P. J. Hakonen, M. A. Sillanpää, Microwave amplification with nanomechanical resonators. *Nature* **480,** 351 – 354 (2011).

23. T. A. Palomaki, J. D. Teufel, R. W. Simmonds, K. W. Lehnert, Entangling mechanical motion with microwave fields. *Science* **342,** 710-713 (2013).

24. J. D. Teufel, D. Li, M. S. Allman, K. Cicak, A. J. Sirois, J. D. Whittaker, R. W. Simmonds, Circuit cavity electromechanics in the strong-coupling regime. *Nature* **471,** 204–208 (2011).

25. Kalaee, M. *et al.* Quantum electromechanics of a hypersonic crystal. *Nat. Nanotechnol.* **14,** 334-339 (2019).

26. Bernier, N. R. *et al*. Nonreciprocal reconfigurable microwave optomechanical circuit. *Nat. Comms.* **8,** 604 (2017).

27. Barzanjeh, S. *et al.* Mechanical on-chip microwave circulator. *Nat. Comms.* **8,** 953 (2017).

28. Kolkowitz, S. *et al.* Coherent sensing of a mechanical resonator with a single-spin qubit. *Science* **335,** 1603-1606 (2012).

29. Teissier, J. *et al*. Strain coupling of a nitrogen-vacancy center spin to a diamond mechanical oscillator. *Phys. Rev. Lett.* **113,** 020503 (2014).

30. Ovartchaiyapong, P., Lee, K. W., Myers, B. A. & Jayich, A. C. B. Dynamic strain-mediated coupling of a single diamond spin to a mechanical resonator. *Nat. Comms.* **5,** 4429 (2014).

31. Li, P. B., Zhou, Y., Gao, W. B. & Nori, F. Enhancing spin-phonon and spin-spin interactions using linear resources in a hybrid quantum system. *Phys. Rev. Lett.* **125,** 153602 (2020).

32. Suh, J. *et al*. Mechanically detecting and avoiding the quantum fluctuations of a microwave field. *Science* **344,** 1262-1265 (2014).

33. Miller, J. M. L. *et al*. Effective quality factor tuning mechanisms in micromechanical resonators. *Appl. Phys. Rev.* **5,** 041307 (2018).

34. Lukin, M. D. & Imamoglu, A. Controlling photons using electromagnetically induced transparency. *Nature* **413,** 273-276 (2001).

35. Kwon, S. *et al.* Magnetic field dependent microwave losses in superconducting niobium microstrip resonators. *J. Appl. Phys.* **124,** 033903 (2018).

36. Degen, C. L., Poggio, M., Mamin, H. J., Rettner, C. T. & Rugar, D. Nanoscale magnetic resonance imaging. *PNAS* **106,** 1313-1317 (2009).

37. Grob, U. *et al.* Magnetic resonance force microscopy with a one-dimensional resolution of 0.9 nanometers. *Nano Lett.* **19,** 7935-7940 (2019).

38. Fischer, R. *et al.* Spin detection with a micromechanical trampoline: towards magnetic resonance microscopy harnessing cavity optomechanics. *New J. Phys.* **21,** 043049 (2019).

39. Rabl, P. *et al*. A quantum spin transducer based on nanoelectromechanical resonator arrays. *Nat. Phys*. **6,** 602-608 (2010).

40. Oeckinghaus, T. *et al.* Spin-phonon interfaces in coupled nanomechanical cantilevers. *Nano Lett.* **20,** 463-469 (2020).

41. Hertzberg, J. B. *et al*. Back-action-evading measurements of nanomechanical motion. *Nat. Phys.* **6,** 213-217 (2010).

**Acknowledgements**

We thank the support from the Fab Infra Team of the Korea Research Institute of Standards and Science. J.C., S.B.S, and J.S. acknowledge support by Korea Research Institute of Standards and Science (KRISS-2020-GP2020-0010). J.S. also acknowledges Samsung Science & Technology Foundation (SSTF-BA1801-03), and the National Research Foundation of Korea (NRF) funded by the Ministry of Science and ICT (2016R1A5A1008184).

**Author contributions**

J.C. and J.S. conceived the idea of the research and designed the experiment. J.C., H.S.K., and J.K contributed to the device fabrication. J.C., S.B.S., and J.S performed the measurements and analyzed the data. J.C. and J.S wrote the manuscript.

**Competing interests**

The authors declare no competing interests.

**Correspondence and requests for materials** should be addressed to J.S.

## Methods

**Sample fabrication.** The device fabrication begins with depositing 100 nm niobium on a sapphire wafer using DC magnetron sputtering with argon gases. We perform photolithography to pattern the input and output coplanar waveguide, the ground plane and the bottom electrode for the nanomechanical resonator. The niobium layer is then removed using and inductively coupled plasma (ICP) etching with $SF_6$ and $O_2$ gases. After removing the photoresist in N-Methyl-2-pyrrolidone (NMP) solvent, we spin-coat 250 nm of 950K A4 Poly-methyl-methacrylate (PMMA) electron-beam resist and define the pattern of the sacrificial layer using electron-beam (e-beam) lithography. We develop the PMMA resist in a mixture of MIBK and IPA. We then smooth the edges of the developed PMMA using the reflow process for 20 minutes at 122 C. A 130 nm thick niobium layer for the nanomechanical resonator and the spiral inductor is deposited using DC sputtering. For the next e-beam exposure, we spin-coat 500 nm ZEP 520 e-beam resist. In this e-beam lithography, we define the spiral inductor and the nanomechanical resonator. The ZEP e-beam resist is then developed using a ZED-N50 solution. Using the ZEP resist as an etch mask, we carry out ICP etching of niobium. We then put our device to N-Methyl-2-pyrrolidone (NMP) solvent to remove the residual ZEP resist and the sacrificial PMMA layer. After this releasing process, we employ a critical point dryer to remove the remaining solvent in our device to prevent adhesion during this drying process.

**Experimental setup.** The fabricated device is diced into a 6 mm x 4mm die and is mounted on a sample holder with two microwave ports with 50 Ohm impedance. The sample is characterized at 4.2 K. The probe and the drive microwave signals are combined at the room temperature using a directional coupler. The probe is attenuated to – 20 dB at the output of the directional coupler. This combined signal is then attenuated at the 4 K stage to – 20 dB and is sent to the input microwave port of the sample holder. The signal transmitted through the device is sent to a microwave circulator to isolator the device from the noise incident to the output port. In the next stage, a high electron mobility transistor (HEMT) amplifier is used to amplify the signal about 40 dB. At the room temperature, a low noise amplifier with 20 dB gain is employed. To characterize the noise spectrum of our device, we connect the output port to a spectrum analyzer without sending the probe signal. For the measurement of microwave transmission, we use a network analyzer to characterize the electromechanically induced reflection.

**Calibration of microwave circuit.** To calibrate the transmission characteristic of our device, we measure a transmission of the measurement circuit with a "thru" reference in place of the device at 4.2 K. Assuming that the contribution of the sample package is negligible, we extract the external decay rate, $\kappa_{ext}$, of our device using the equation, $\kappa_{ext} = 10^{[S21(\omega=\omega_c) - S21,0(\omega=\omega_c)]/20} \kappa$. Here, $S_{21,0}(\omega = \omega_c)$ is the transmission signal with the thru reference and $\kappa$ is the total decay rate. The intrinsic decay rate, $\kappa_o$ is $\kappa_o = \kappa - \kappa_{ext}$.

**Relation of the sideband power and the displacement of the mechanical resonator.** We derive the relation of the sideband power and the displacement of the nanomechanical resonator by modeling our device using circuit elements (Fig. 1d). To simplify the analysis, we covert the circuit in Figure 1d into an equivalent circuit consisting of parallel components only (Supplementary Fig. 3). We solve for the voltage in the circuit, $V(t)$. The microwave drive signal is modelled using a current source, $I(t)$, with amplitude $I_0$ and frequency $\omega_d$. Using Kirchhoff's current law, the current flowing in the circuit is $I_0 \cos \omega_d t = \frac{\partial [C_T V(t)]}{\partial t} + R_T^{-1} V(t) + L^{-1} \int V(t)\, dt$, where $C_T = C_m(x) + C_0 + 2C_{\kappa,eq}$ and $R_T = (R_0^{-1} + 2R_{L,eq}^{-1})^{-1}$ are the total capacitance and the total resistance of the circuit. Here, $C_m(x)$ is the capacitance of the nanomechanical resonator, $C_0$ is the capacitance of other parasitic components in the microwave resonator, $C_{\kappa,eq}$ is the equivalent capacitance for the coupling capacitor, $R_0$ is the intrinsic resistance of the circuit, and $R_{L,eq}$ is the equivalent resistance for the load resistance. We approximate $C_m(x)$ up to the first order and express it as

$$C_m(x) = C_{m,x=0} + \frac{\partial C_m}{\partial x} x_0 \cos(\Omega_m t + \phi_m).$$

Now $C_T(x) = C_{T,x=0} + \frac{\partial C_m}{\partial x} x_0 \cos(\Omega_m t + \phi_m)$, where $C_{T,x=0} = C_{m,x=0} + C_0 + 2C_{\kappa,eq}$. If we let $C_{m,x=0} + C_0 = C$, multiply $\sqrt{L/C}$ to the current equation, differentiate once and organize the equation in terms of $V$, $\partial V/\partial t$ and $\partial^2 V/\partial t^2$, we obtain

$$-I_0\sqrt{L/C}\,\omega_d \sin(\omega_d t)$$

$$= V\left(\frac{1}{\sqrt{LC}} - \sqrt{\frac{L}{C}}\,\omega_m^2 \frac{\partial C_m}{\partial x} x_0 \cos(\Omega_m t + \phi_m)\right)$$

$$+ \frac{\partial V}{\partial t}\left(\frac{1}{R_T}\sqrt{\frac{L}{C}} - 2\sqrt{\frac{L}{C}}\,\omega_m \frac{\partial C_m}{\partial x} \sin(\Omega_m t + \phi_m)\right)$$

$$+ \frac{\partial^2 V}{\partial t^2}\left(\sqrt{\frac{L}{C}} C_{T,x=0} + \sqrt{\frac{L}{C}} \frac{\partial C_m}{\partial x} x_0 \cos(\Omega_m t + \phi_m)\right).$$

Assume that the coupling capacitance is negligible, we get $C_{T,x=0} = C$. Also, the total decay rate of the microwave resonator is $\kappa = (R_T C)^{-1}$ and the microwave resonant frequency is $\omega_c = (LC)^{-\frac{1}{2}}$. Using these definitions, the equation is modified to

$$-I_0\sqrt{L/C}\,\omega_d \sin(\omega_d t)$$

$$= \omega_c V\left(1 - \frac{1}{C}\frac{\omega_m^2}{\omega_c^2}\frac{\partial C_m}{\partial x} x_0 \cos(\Omega_m t + \phi_m)\right)$$

$$+ \frac{\kappa}{\omega_c}\frac{\partial V}{\partial t}\left(1 - \frac{2}{C}\frac{\omega_m}{\kappa}\frac{\partial C_m}{\partial x} x_0 \sin(\Omega_m t + \phi_m)\right)$$

$$+ \frac{1}{\omega_c}\frac{\partial^2 V}{\partial t^2}\left(1 + \frac{1}{C}\frac{\partial C_m}{\partial x} x_0 \cos(\Omega_m t + \phi_m)\right).$$

As we are interested in the voltage signal of anti-Stokes sideband (sum frequency, $\omega_s = \omega_d + \omega_m$) created by the microwave drive at $\omega_d = \omega_c - \Omega_m$, it is reasonable to assume the voltage solution as $V(t) = V_d \cos(\omega_d t + \phi_d) + V_s \cos(\omega_s t + \phi_s)$. If we insert this solution to the equation above and solve for $V_s^2/V_d^2$ by assuming that any terms of order $\Omega_m/\omega_c$ and $\Omega_m^2/\omega_c^2$ are negligible, $\omega_s/\omega_c = \omega_d/\omega_c \approx 1$ and $\omega_c - \omega_s^2/\omega_c \approx 2(\omega_c - \omega_s)$, we obtain a solution for the sideband voltage at $\omega_s = \omega_c$, in terms of the microwave drive voltage. The equation is given by

$$\frac{P_{sideband}}{P_d} = \frac{V_s^2}{V_p^2} = \frac{\left(\frac{\partial \omega_c}{\partial x}\right)^2}{\kappa} x_0^2 = \frac{2g_0^2}{\kappa^2 x_{zpf}^2}\langle x^2(t)\rangle.$$

Here, $P_{sideband}$ is the sideband power and $P_d$ is the driving power, both measured at the spectrum analyzer. Note we use the following relation:

$$\frac{1}{2C}\frac{\partial C_m}{\partial x} = -\frac{1}{\omega_c}\frac{\partial \omega_c}{\partial x}, \left(\frac{\partial \omega_c}{\partial x}\right)^2 = \frac{g_0^2}{x_{zpf}^2}, \text{and } x_0^2 = 2\langle x^2(t)\rangle.$$

**Calculation of single-photon electromechanical coupling rate $g_0$.** The noise spectrum of a thermally fluctuating nanomechanical resonator with eigenfrequency $\omega_m$ and intrinsic damping rate $\Gamma_m$ in a thermal bath of temperature $T_{\text{bath}}$ is given by $S_x(\omega) = \int_{-\infty}^{\infty}\langle x(t)x(0)\rangle e^{i\omega t}dt$, where $x(t)$ is the displacement of the resonator. In the classical regime where $k_B T \gg \hbar\omega_m$, the fluctuation-dissipation theorem gives $S_x(\omega) = \frac{2k_B T}{\omega}\text{Im}\chi_m(\omega)$, where $\chi_m(\omega) = \left[m_{eff}(\Omega_m^2 - \omega^2) - im_{eff}\Gamma_m\omega\right]^{-1}$ is the mechanical susceptibility of the resonator. The variance of the displacement is related to the energy of the resonator in thermal equilibrium and is obtained by integrating the area of $S_x(\omega)$. This is expressed as $\langle x^2(t)\rangle = \int_{-\infty}^{\infty} S_x(\omega)d\omega/2\pi = \frac{k_B T}{m_{eff}\Omega_m^2}$, according to the equipartition theorem.

To obtain the single-photon optomechanical coupling rate, $g_0$, we weakly drive the microwave resonator in the red-detuned regime ($\Delta = -\Omega_m$) to avoid optomechancial back-action effects. We then measure the power of the up-converted sideband signal, $P_{sideband}$, around $\omega_c$. We extract $g_0$ from the normalized sideband power at $T = 4.17$ K using the equation we derive in the previous section,

$$\frac{P_{sideband}}{P_d} = \frac{2g_0^2}{\kappa^2 x_{zpf}^2}\langle x^2(t)\rangle = \frac{2g_0^2}{\kappa^2 x_{zpf}^2}\frac{k_B T}{m_{eff}\Omega_m^2} = \frac{4g_0^2}{\kappa^2}\frac{k_B T}{\hbar\Omega_m}.$$

Here, $P_d$ is the power of the microwave drive measured at the spectrum analyzer.

**Calibration of phonon and drive photon numbers.** To calibrate the drive photon numbers, $n_d$, induced in the cavity, we measure $P_{sideband}$ around $\omega_c$ for different $P_d$. We then extract the total mechanical damping rate, $\Gamma_T = \Gamma_m + \Gamma_{opt}$ from the Lorentzian fitting of the sideband signals. Note that $\Gamma_{opt}$ is linearly dependent on the drive power. If we let $\Gamma_{opt} = \beta P_d$ and equate this with $\Gamma_{opt} = 4g_0^2 n_d/\kappa$, we get $\frac{n_d}{P_d} = \frac{\kappa}{4g^2}\beta$, where $\beta$ is obtained by fitting the experimental $\Gamma_T$ with respect to $P_d$.

The phonon number, $n_{ph}$, can be calculated considering the normalized sideband power, $\frac{P_{sideband}}{P_d} = \frac{4g_0^2}{\kappa^2}\frac{k_B T_{bath}}{\hbar\Omega_m}$. At a high temperature ($k_B T_{bath} \gg \hbar\Omega_m$), the occupation factor is $n_{ph} = [\exp(\hbar\Omega_m/k_B T_{bath}) - 1]^{-1} \approx k_B T_{bath}/\hbar\Omega_m$. By measuring $\frac{P_{sideband}}{P_d}$ for different $n_d$, the phonon occupation can be obtained from $n_{ph}(n_d) = \frac{P_{sideband}}{P_d}\left(\frac{\kappa^2}{4g_0^2}\right)$.

The application of the external magnetic field affects the quality of the microwave resonator, increasing its decay rate, $\kappa$. Assuming the microwave resonator exhibits Lorentzian resonance curve, the equation for the drive photon number is given by $n_d = \frac{1}{\hbar\omega_d}\frac{2}{\kappa_{ext}}G\frac{\kappa_{ext}^2}{\kappa^2+4(\omega_d-\omega_c)^2}P_{in}$, where $G$ is the total gain of the microwave circuit, $P_{in}$ is the input power from the microwave source generator[41]. If we assume that $G$ and $\kappa_{ext}$ do not change with the external magnetic field, it is possible to calculate $n_d$ for $B > 0$ using the following equation: $n_d = \frac{\omega_d(B=0)}{\omega_d}\frac{\kappa^2(B=0)+4[\omega_d(B=0)-\omega_c(B=0)]^2}{\kappa^2(B)+4(\omega_d(B)-\omega_c(B))^2}n_d(B=0)$.

**Calculation of displacement and force noise spectra from noise power spectrum.**
The total displacement noise spectrum, $S_x(\omega)$, near the microwave resonance can be expressed as the sum of the mechanical sideband noise spectrum, $S_x^{mech}(\omega)$, and the background noise spectrum, $S_x^{bg}(\omega)$. This background noise is usually a white noise originates from the measurement circuits and thermal fluctuations in the microwave resonator. To begin with the derivation, we consider the normalized sideband power, which is given by

$$\frac{P_{sideband}}{P_d} = \frac{2g_0^2}{\kappa^2 x_{zpf}^2}\langle x^2(t)\rangle.$$

The variance of the mechanical displacement is $\langle x^2(t)\rangle = \int_{-\infty}^{\infty}S_x^{mech}(\omega)d\omega/2\pi$. $P_{sideband}$ is the power contained in mechanical noise power spectrum $S_v^{mech}(\omega)$ and is defined as $P_{sideband} = \int_{-\infty}^{\infty}S_v^{mech}(\omega)d\omega/2\pi$. Using the relations and $P_d^{-1} = P_m^{-1}(4g_0^2/\kappa^2)(k_B T_{bath}/\hbar\Omega_m)$, we obtain

$$S_x^{mech}(\omega) = \frac{S_v^{mech}(\omega)}{P_d}\frac{\kappa^2 x_{zpf}^2}{2g_0^2} = \frac{S_v^{mech}(\omega)}{P_m}\frac{\kappa^2 x_{zpf}^2}{2g_0^2}\frac{4g_0^2}{\kappa^2}\frac{k_B T_{bath}}{\hbar\Omega_m} = \frac{S_v^{mech}(\omega)}{P_m}\frac{k_B T_{bath}}{k}.$$

Using this equation, we calculate the background noise spectrum as

$$S_x^{bg}(\omega) = \frac{S_v^{bg}(\omega)}{P_m}\frac{k_B T_{bath}}{k}.$$

The force noises corresponding to $S_f^{mech}(\omega)$ and $S_f^{bg}(\omega)$ are calculated by dividing $S_x^{mech}(\omega)$ and $S_x^{bg}(\omega)$ by the absolute square of the mechanical susceptibility, $\chi_m(\omega) = \left[m_{eff}(\Omega_m^2 - \omega^2) - im_{eff}\Gamma_m\omega\right]^{-1}$. The force sensitivity and the background noise at the mechanical resonant frequency are therefore given by

$$S_f^{mech}(\Omega_m) = \frac{S_x^{mech}(\Omega_m)}{|\chi_m(\Omega_m)|^2} \text{ and } S_f^{bg}(\Omega_m) = \frac{S_x^{bg}(\Omega_m)}{|\chi_m(\Omega_m)|^2}.$$

# Supplementary Information

# Superconducting Nanoelectromechanical Transducer Resilient to Magnetic Fields

Jinwoong Cha[1], Hak-Seong Kim[1], Jihwan Kim[1,2], Seung-Bo Shim[1], and Junho Suh[1,*]

[1]Quantum Technology Institute, Korea Research Institute of Standards and Science, Daejeon, South Korea

[2]Department of Physics, Korea Advanced Institute of Science and Technology, Daejeon, South Korea

Correspondence to Junho Suh: junho.suh@kriss.re.kr

**Supplementary Figure 1.** Electromechanical back-action effects.

**Supplementary Figure 2.** Signal-to-noise ratio (SNR) and force sensitivity of the niobium device.

**Supplementary Figure 3.** Equivalent circuit model for the niobium device.

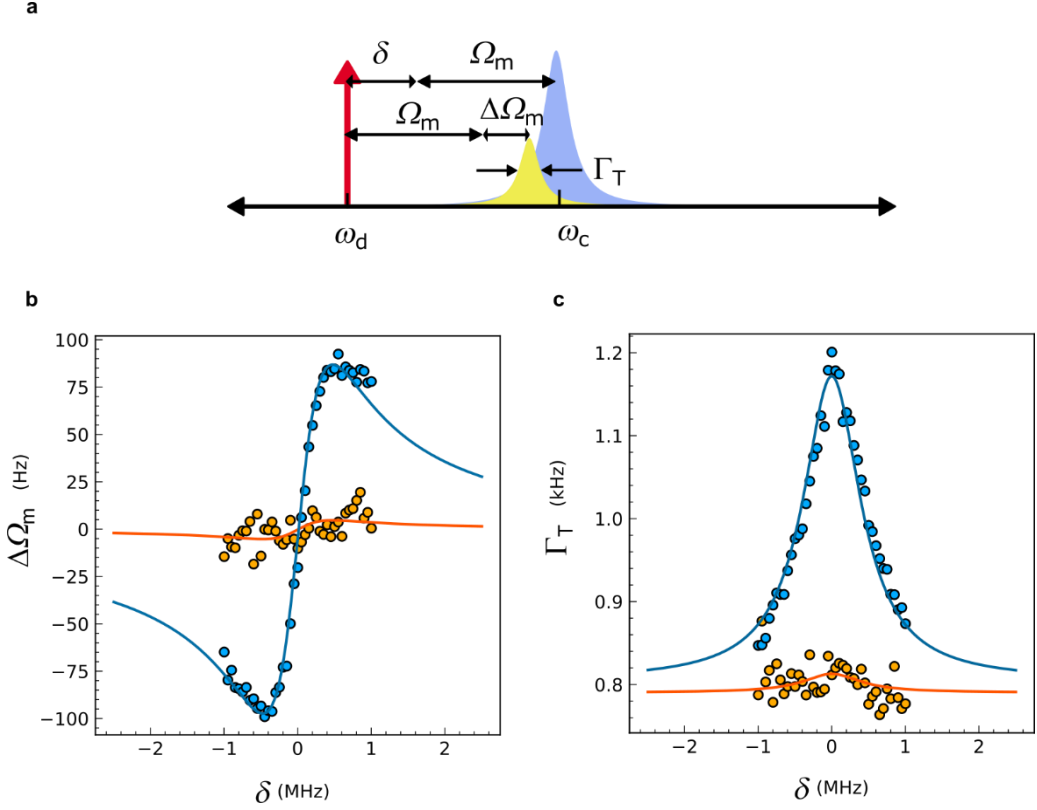

**Supplementary Figure. 1 Electromechanical back-action effects. a** Schematic of the experiment in frequency domain. The frequency of the red-detuned microwave drive, $\omega_d$, is equal to $(\omega_c - \Omega_m) + \delta$ and $\delta$ changes from −1 MHz to 1 MHz. This drive signal interacting with the mechanical resonator leads to the sideband signal (yellow Lorentzian curve) at $\omega_d + \Omega_m + \Delta\Omega_m$. Here, $\Omega_m$ is the resonant frequency of the mechanical resonator when $\delta = 0$. $\Delta\Omega_m$ is the shift of the mechanical resonant frequency due to the electromechanical spring effect. $\Gamma_T$ is the linewidth of the mechanical resonator. **b** $\Delta\Omega_m$ as a function of $\delta$. **c** $\Gamma_T$ as a function of $\delta$. The blue (orange) circles in (**b**) and (**c**) represent experimentally obtained data when the microwave drive power is $n_d \approx 8.1 \times 10^6$ ($n_d \approx 2.0 \times 10^5$). The solid lines in (**b**) and (**c**) show the fitting to the equations for (**b**) $\Delta\Omega_m \left(= \Omega_m + 4g_0^2 n_d \left[\frac{\Delta+\Omega_m}{\kappa^2+4(\Delta+\Omega_m)^2} + \frac{\Delta-\Omega_m}{\kappa^2+4(\Delta-\Omega_m)^2}\right]\right.$ and (**c**) $\Gamma_T \left(= \Gamma_m + 4g_0^2 n_d \left[\frac{\kappa}{\kappa^2+4(\Delta+\Omega_m)^2} + \frac{\kappa}{\kappa^2+4(\Delta-\Omega_m)^2}\right]\right)$.

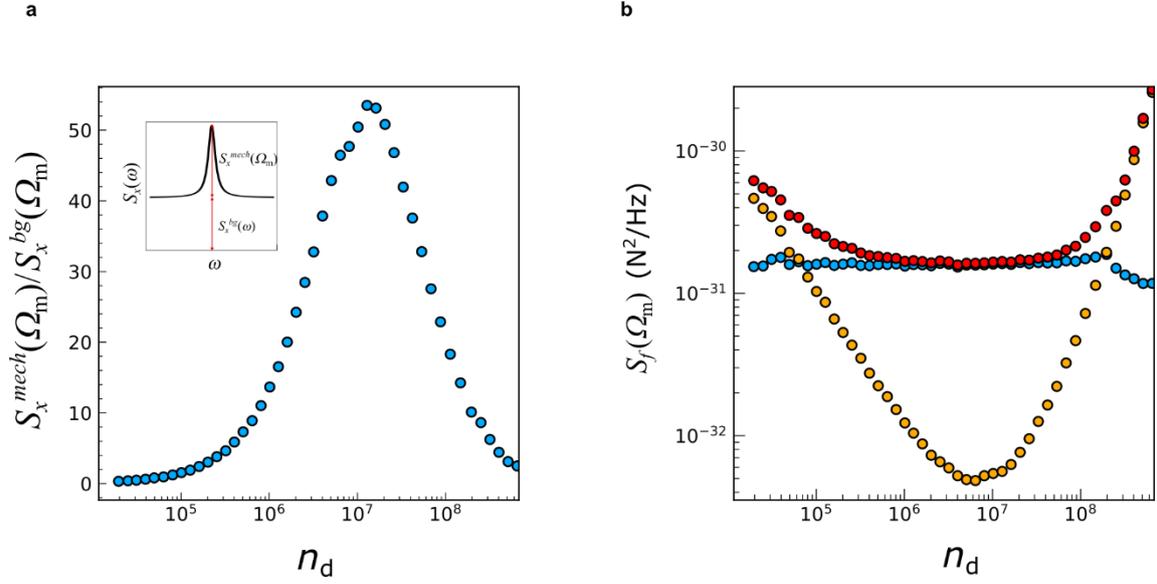

**Supplementary Figure. 2 Signal-to-noise ratio (SNR) and force sensitivity of the niobium device. a** SNR as a function of the pump power $n_d$. SNR is defined as $S_x^{mech}(\Omega_m)/S_x^{bg}(\Omega_m)$. $S_x^{mech}(\Omega_m)$ is the height of the mechanical displacement noise and the $S_x^{bg}(\Omega_m)$ is the height of the background displacement noise. **b** The force sensitivity of the device as a function of $n_d$. The blue, orange, and red solid circles correspond to $S_f^{mech}(\Omega_m)$, $S_f^{bg}(\Omega_m)$, and $S_f^{mech}(\Omega_m) + S_f^{bg}(\Omega_m)$, respectively. The procedure for the conversion between the displacement noise and the force noise is described in Methods.

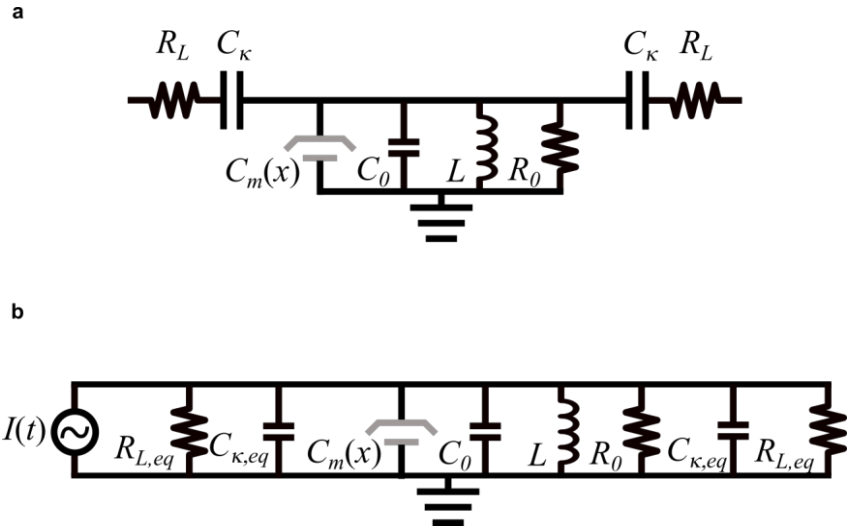

**Supplementary Figure. 3 Equivalent circuit model for the niobium device. a** Circuit with both serial and parallel components. $R_L$ is the load resistance, $C_\kappa$ is the coupling capacitance, $C_m$ is the capacitance of the nanomechanical resonator, $C_0$ is the capacitance originating from parasitic components, $L$ is the inductance of the microwave resonator, $R_0$ is the resistance from intrinsic lossy components. **b** Parallel circuit equivalent to the circuit in (**a**). Here, all the coupling components are converted into equivalent parallel components $R_{L,eq}$ and $C_{\kappa,eq}$. The current source is the Norton equivalent current.